\documentclass{iaus}
\usepackage{graphicx}

\title[Age spreads in SFRs?] 
{Age spreads in star forming regions?}

\author[R. D. Jeffries]   
{R. D. Jeffries}

\affiliation{Astrophysics Group, Keele University, Keele,
  Staffordshire, ST5 5BG, UK\\
email: {\tt rdj@astro.keele.ac.uk}}

\pubyear{2008}
\volume{258} 
\pagerange{1--6}
\jname{The Ages of Stars}
\editors{E.E. Mamajek, D.R. Soderblom \& R.F.G. Wyse, eds.}
\begin{document}

\maketitle

\begin{abstract}
Rotation periods and projected equatorial velocities of
pre-main-sequence (PMS) stars in star forming regions can be combined
to give projected stellar radii. Assuming random axial orientation, a
Monte-Carlo model is used to illustrate that distributions of projected
stellar radii are very sensitive to ages and age dispersions between 1
and 10\,Myr which, unlike age estimates from conventional
Hertzsprung-Russell diagrams, are relatively immune to uncertainties
due to extinction, variability, distance etc. Application of the
technique to the Orion Nebula cluster reveals radius spreads of a
factor of 2--3 (FWHM) at a given effective temperature. Modelling this
dispersion as an age spread suggests that PMS stars in the ONC have an
age range larger than the mean cluster age, that could be reasonably
described by the age distribution deduced from the
Hertzsprung-Russell diagram. These radius/age spreads are certainly
large enough to invalidate the assumption of coevality when considering
the evolution of PMS properties (rotation, disks etc.) from one young
cluster to another.

\end{abstract}

\firstsection 
\section{Introduction}
Does star formation take a long time, or is it all over on a dynamical
free-fall timescale? This is a keenly debated question
in star formation theory, with implications spanning topics as diverse
as investigating early star/disk/planet evolution using populations in
young star formation regions (SFRs) which are often {\it assumed to be
coeval}, through to assessing overall star formation efficiency and the
build up of galactic populations.

According to one paradigm, the collapse of molecular clouds is a
quasi-static process slowed by magnetic pressure. The timescale for
star formation is governed by ambipolar diffusion and could be $\simeq
10$\,Myr (e.g. Tan, Krumholz \& McKee 2006).  Alternatively, on the
basis of short deduced molecular cloud lifetimes, others argue that
star formation is a rapid process, taking place in compressed
filamentary structures on free-fall timescales $\leq 1$\,Myr (e.g.
Elmegreen 2007).

A crucial piece of evidence for star formation timescales is the
presence (or not) of age spreads among stars in young SFRs. Low-mass
pre-main-sequence (PMS) stars can be assigned model-dependent ages from
their position in Hertzsprung-Russell (H-R) diagrams as they contract
along Hayashi tracks. Using this technique several authors (e.g. Palla
\& Stahler 2000; Huff \& Stahler 2006) 
claim star formation ``accelerates'' exponentially up to the
present day, on timescales of $\simeq 10$\,Myr.  These apparent
age spreads favour quasi-static, ``slow'' star formation. However,
conventional H-R diagrams are severely affected by (i) intrinsic
variability, (ii) extinction uncertainties, (iii) accretion luminosity,
(iv) binarity, (v) distance dispersion -- all of which can mimic age
spreads where none exist (e.g. Hartmann 2001; Hillenbrand, Bauermeister
\& White 2008).

In this contribution I illustrate a technique to circumvent these difficulties
using the rotational properties of PMS stars. This produces an
alternative H-R diagram (radius versus temperature) that
can be modelled to reconstruct a star formation history
free from the problems above (e.g. see Jeffries 2007a,b).

\section{Projected stellar radii}

New wide-field surveys are finding rotation periods ($P$ in days) for
hundreds of magnetically spotted PMS stars in SFRs. At the same time, it
is now possible to obtain projected equatorial velocities ($v \sin i$
in km\,s$^{-1}$) for these stars from rotational line
broadening using multi-object spectrographs such as FLAMES at the VLT
and Hectoechelle at the MMT.  Combining these measurements gives
geometric estimates of radii, $R \sin i = 0.02\, P\, v \sin i$ (in
solar radii). The inclination angle, $i$, is unknown, but if it is
assumed random (for which there is some evidence -- e.g. Jackson \&
Jeffries in these proceedings -- and no counter evidence) and the
measurement uncertainties are understood, then distributions of $R \sin
i$ can be Monte-Carlo modelled to estimate the true $R$ for any group
of stars.

As an example of the technique's power, in Fig.~1 I show a {\it
  simulation} of what could be achieved by observing projected
  equatorial velocities for 458 PMS objects with rotation periods in
  the young SFR NGC~2264 
  (from Lamm et al. 2004 and Makidon et al. 2004), with a 10\%
  precision and a threshold for detection of $v \sin i \geq
  15$\,km\,s$^{-1}$ -- which is routinely possible. The simulation assumes
  that rotation axes are randomly oriented but that objects with
  $i<30^{\circ}$ do not show rotational modulation. The left hand
  panels show the recovered $R\sin i$ values versus $V-I$ for coeval
  populations at several ages, where the 
  Siess, Dufour \& Forestini (2000, S00) isochrones
  are used to assign the intrinsic stellar radii. The right hand panels
  collapse this distribution to 1-dimensional form by normalising
  $R\sin i$ at each colour by the value of $R$ at 3\,Myr.

With typical measurements, a set of 20 $R\sin i$ values can give $R$ to
$\pm 5\%$. But at a given colour or $T_{\rm eff}$, $R$ is expected to
change by a factor of three between 1 and 10\,Myr! Hence the $R\sin i$
distribution is {\it very sensitive} to age differences and age
dispersions in this range (see right panels of Fig.~1), 
but becomes less so at older
ages.  Any inferred ages and age spreads are of course
model-dependent, but the radii are absolute.  The technique is almost
immune to problems associated with variability, binarity, extinction
uncertainty and accretion luminosity. It is also distance-independent
to boot!

\begin{figure}
\begin{center}
 \includegraphics[width=4.7in]{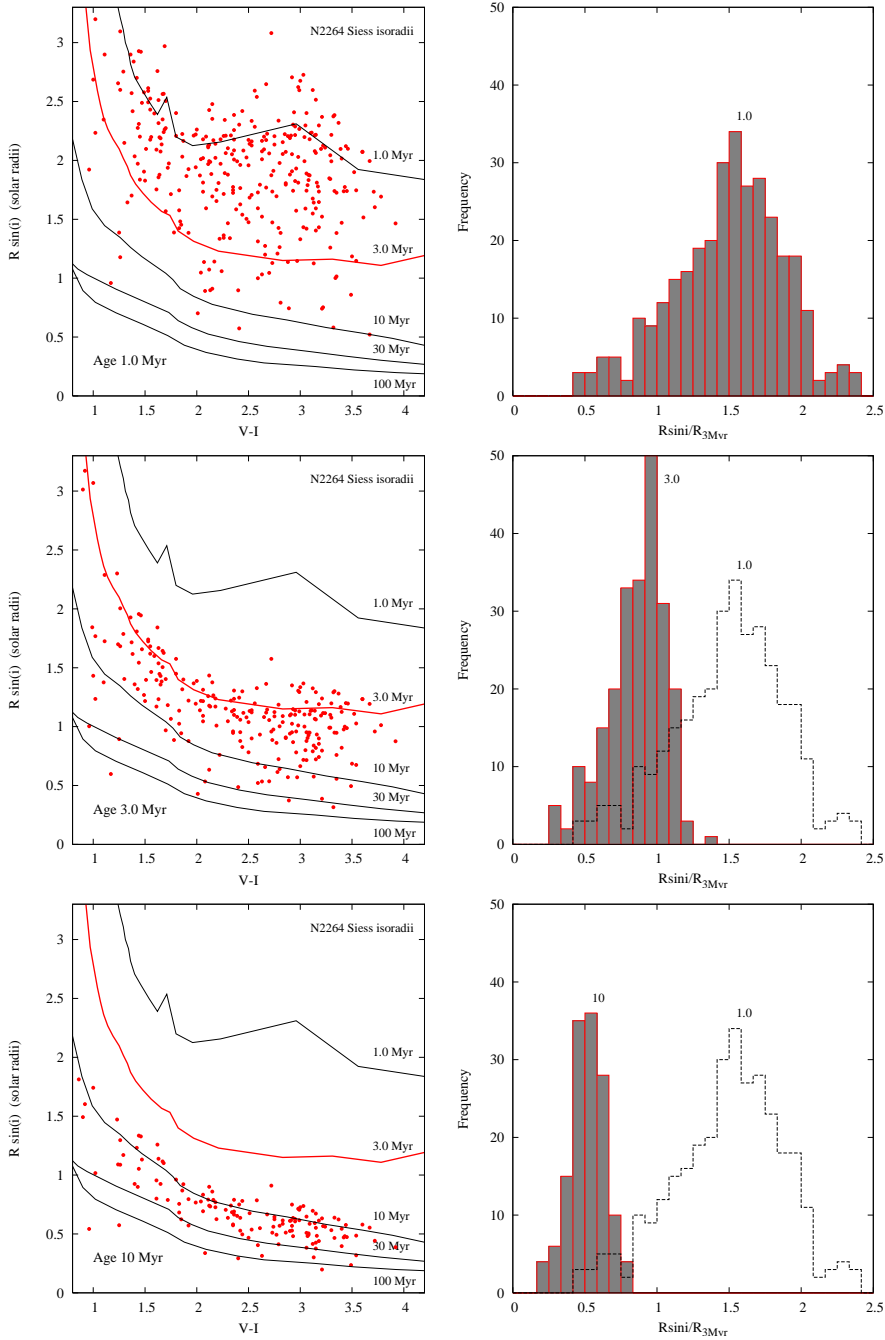} 
 \caption{
A simulation of the expected $R\sin i$ values that would be obtained
from a sample of 458 periodic PMS stars in the young SFR NGC~2264.
The left hand panels show $R\sin i$ values versus colour assuming
random rotation axis orientation and that only $v \sin i$ values $\geq
15$\,km\,s$^{-1}$ are detectable. The right hand panels show the
1-dimensional collapsed distribution obtained by normalising by the
expected radius at an age of 3\,Myr. The simulations include typical
measurement uncertainties in colour, $v\sin i$ and period. Each row
shows how the distribution would look if the NGC~2264 stars were coeval
and at ages of 1, 3 or 10\,Myr. The solid lines are radius isochrones from
Siess et al. (2000).
}
   \label{fig1}
\end{center}
\end{figure}

\section{Results for the Orion Nebula Cluster}

\begin{figure}
\begin{center}
 \includegraphics[width=5.2in]{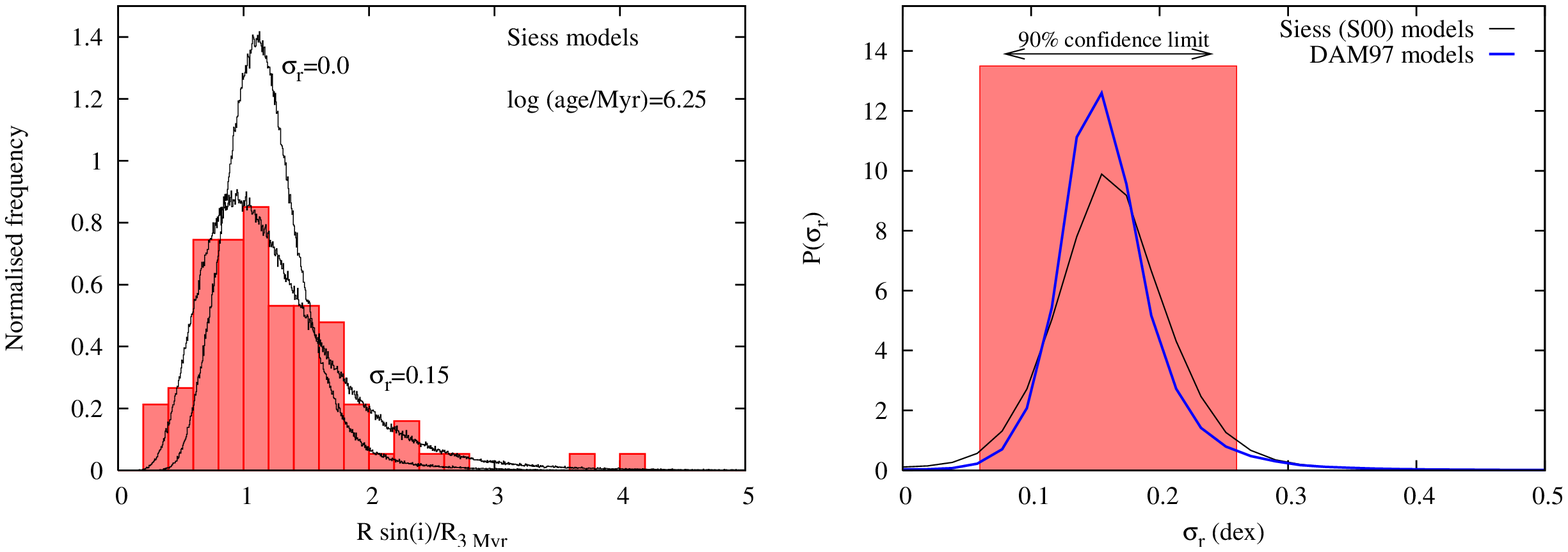} 
 \caption{(Left) The measured normalised $R \sin i$ distribution for the
 ONC compared with a coeval model ($\sigma_r=0.0$). This model
 distribution is too narrow. Also shown is a model with a
 Gaussian spread ($\sigma_r = 0.15$\,dex) in $\log_{10} R$ which
 provides a much better fit. In both
 cases, the central age for the distribution is 1.78\,Myr. (Right) The
 probability distribution for $\sigma_r$ using either the S00 or DAM97
 isochrones. The 90\% confidence region for the S00 isochrones is
 shown, but the result is almost identical for the DAM97 isochrones.
 This modelling implies a spread of a factor 2--3 (FWHM) in radius at a
 given $T_{\rm eff}$.
}
   \label{fig2}
\end{center}
\end{figure}

The first attempts to use this technique were made in the Orion Nebula
Cluster (ONC). The results were described in detail by Jeffries
(2007b) and are summarised here. The ONC is a young and populous SFR
with a sample of 95 K- and M-stars that have measured rotation periods
(Herbst et al. 2002), effective temperatures (Hillenbrand 1997) and
$v\sin i$ (Rhode, Herbst \& Mathieu 2001; Sicilia-Aguilar et al. 2005).

I calculated $R\sin i$ for these stars and then simulated the
normalised (to $R_{\rm 3Myr}$) distributions using the Monte Carlo
model which produced the simulations in Fig.~1. The models were tested
against the observed data using the Kolmogorov-Smirnoff statistic on
the cumulative distributions.  The simulations are insensitive to the
threshold $i$ below which it is assumed no rotational periodicity would
be found, but {\it are} sensitive to the choice of radius
isochrones. I ran models using the S00 and
D'Antona \& Mazzitelli (1997, DAM97) isochrones. Uncertainties in
periods, $v\sin i$ and $T_{\rm eff}$ were taken from the sources cited
above.

The first models I tried were coeval with the age as a free parameter.
The best-fitting ages were 1.78\,Myr and 0.76\,Myr for the S00 and
DAM97 isochrones, but both were rejected as good models at the $>95\%$
level (see Fig.~2). New models were generated by allowing the radius to
spread around a single coeval value. The spread was characterised by a
Gaussian $\sigma_r$ in $\log_{10} R$. These generated good fits with
central ages very similar to the previous coeval model, but with
$\sigma_r \simeq (0.15 \pm 0.08)$\,dex (90\% confidence interval) for both
sets of isochrones (see Fig.~2). This implies linear radius spreads of
a factor of 2--3 (FWHM) at a given $T_{\rm eff}$.

Rather than a simple radius spread it is natural to interpret the
results in terms of an age spread. I fitted two types of analytic age spread: a
Gaussian spread ($\sigma_a$) in $\log_{10}$\,age about a central value
\[ f(\log_{10}{\rm age}) = N_0 \exp\left(\frac{-(\log_{10}{\rm age}
    - \log_{10}{\rm central\ age})^2}{2\sigma_{a}^{2}}\right)\, ; \]  
or an exponentially accelerating star forming rate with timescale
$\lambda_a$ and an abrupt cut-off (or zero-point) age
\[ f({\rm age}) = N_0 \exp\left( -\frac{{\rm age}}{\lambda_a} \right ) \
\ \ \ {\rm for\ age}>{\rm zeropoint\ age}\, . \]
Finally, I modelled the $R\sin i$
distribution by assuming that the stars had the age distribution
implied by their positions in the H-R diagram (assuming
an ONC distance of 392\,pc -- Jeffries 2007a). 

\begin{figure}
\begin{center}
 \includegraphics[width=5.1in]{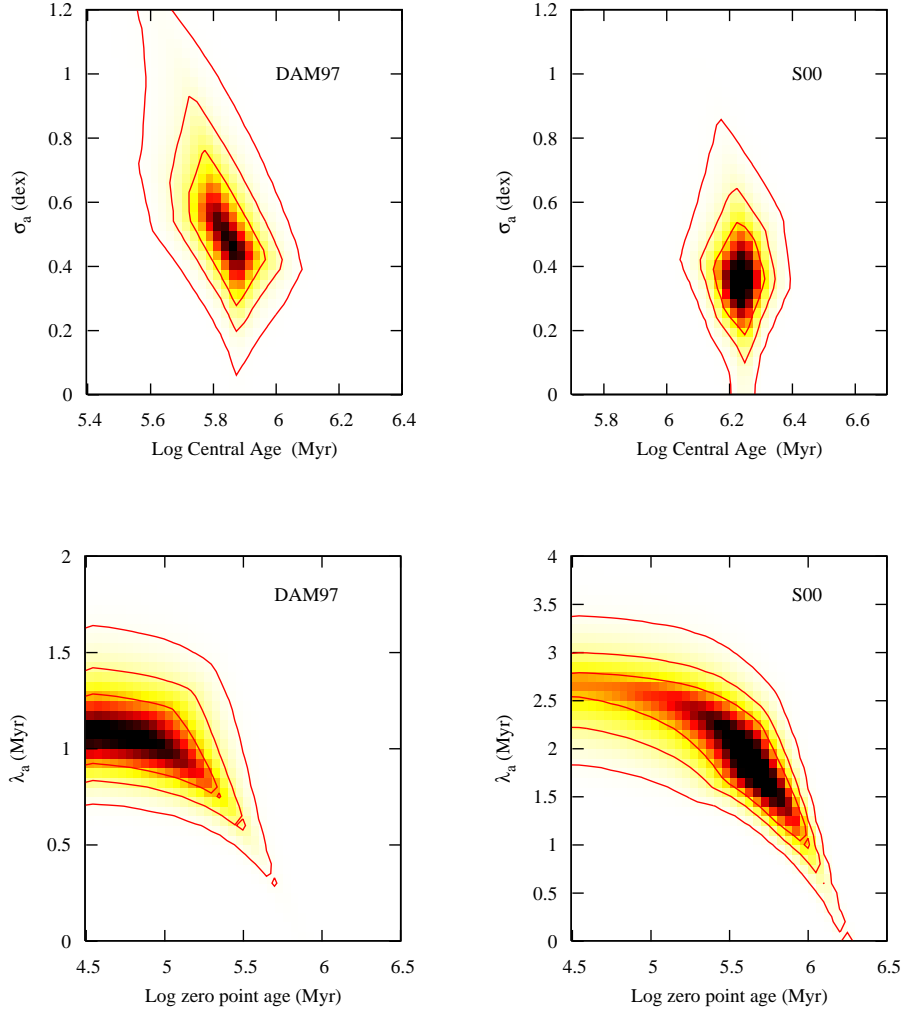} 
 \caption{Relative probability distributions of a good fit in the cases
 of a Gaussian distribution in $\log_{10}$age (top row, with free
 parameters of a central age and dispersion $\sigma_a$ in dex) and an
 exponentially accelerating star formation rate (bottom row, with free
 parameters of a timescale $\lambda_a$ and a zero-point cut-off age. For
 each model we show the results using either the S00 or DAM97
 isochrones. The contours enclose 68\%, 90\% and 99\% of the probability.
}
   \label{fig3}
\end{center}
\end{figure}

\begin{figure}
\begin{center}
 \includegraphics[width=5.2in]{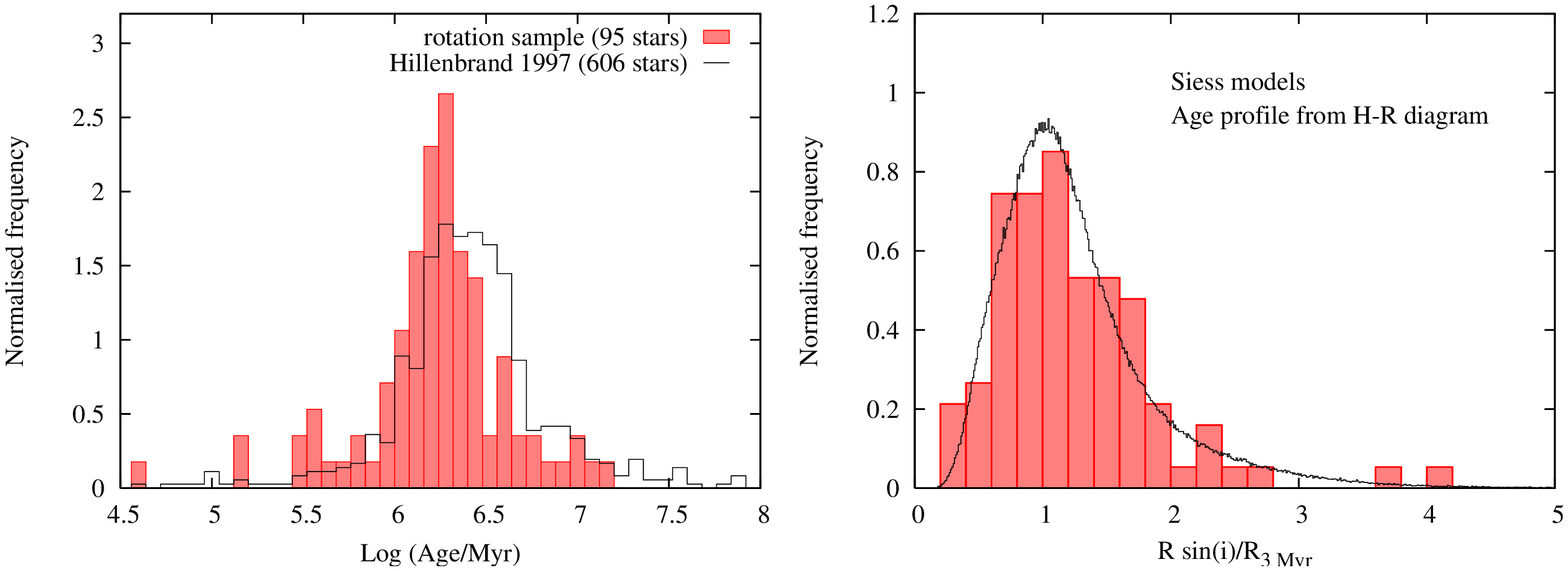} 
 \caption{(Left) Normalised age distribution of ONC PMS stars estimated from
   the H-R diagram and S00 isochrones. I show normalised distributions
   for the rotation sample and the full sample of Hillenbrand
   (1997). The rotation sample is missing some of the ``oldest''
   stars. (Right) The $R\sin i$ distribution modelled using the age
   distribution for the rotation sample in the left-hand panel is a
   reasonable fit to the observed $R\sin i$ distribution.
}
   \label{fig4}
\end{center}
\end{figure}

There are three main results, summarised below.
\begin{enumerate}
\item Both classes of model require an age spread ($\sigma_a > 0$,
  $\lambda_a >0$ -- see Fig.3). For the Gaussian model the best fitting
  dispersion $\sigma_a \simeq 0.4$\,dex is independent of isochrone
  choice, but with model-dependent central ages similar to those given by the
  coeval models. The exponential model has a best-fitting $\lambda_a
  \simeq 1.1$\,Myr for the DAM97 isochrones and $\lambda_a \simeq
  1.9$\,Myr for the S00 isochrones.

\item The data are incapable of distinguishing between the
  exponentially accelerating model or the Gaussian spread in
  $\log_{10}$age.

\item Modelling the $R\sin i$ distribution using the ages derived from
  the H-R diagram (see Fig.~4) gives a reasonable fit for both sets of
  isochrones.
\end{enumerate}

\section{Discussion}

Although the absolute ages and age dispersions derived with this
technique are to some extent
model-dependent, the absolute radii and radius dispersion are geometric
estimates. We conclude that there is very strong evidence for
spreads amounting to factors of 2-3 (FWHM) in radius at a given $T_{\rm
  eff}$ in PMS stars of the ONC. As PMS tracks are close-to-vertical in
the H-R diagram for low-mass stars, 
this implies order-of-magnitude spreads in moment of
inertia -- a fact that cannot be ignored when considering the angular
momentum evolution of PMS stars in SFRs.

Whether these radius spreads represent real age spreads is a moot
point. It is possible that differing accretion histories could lead to
luminosity/radius differences for coeval stars of similar present-day
$T_{\rm eff}$. However, according to current, non-accreting models, the
data imply age spreads in the ONC that are larger than its mean age
($>2$\,Myr for the S00 models), consistent with age spreads
judged from its conventional H-R diagram, and certainly large enough to
compromise any coeval assumption. In addition, the spreads we
have found may actually be underestimates.  The rotation sample in the
ONC is clearly biased against the faintest (possibly oldest?)  stars
(see Fig.~4).  We cannot comment on whether age spreads as large as
10\,Myr are likely until these low-luminosity outliers in the ONC have
their periods and projected rotation velocities measured.

\end{document}